\documentclass[preprint,12pt]{elsarticle}
\usepackage{epsfig}
\usepackage{graphics}
\usepackage{amssymb}
\usepackage{lipsum}
\usepackage{textcomp}
\usepackage{amsmath}
\usepackage[dvipsnames]{xcolor}
\usepackage{subfigure}
\usepackage{braket}
\usepackage{tikz}
\usetikzlibrary{arrows,decorations.markings}

\def\eq#1{(\ref{#1})}

\journal{Nuclear Physics B}

\begin{document}

\title{A note on the AdS/CFT correspondence and the nature of spacetime in quantum gravity.}

\author{Carlos Silva} 

\affiliation{organization={Instituto Federal de Educação Ciência e Tecnologia do Ceará (IFCE),  Campus Tianguá},
            addressline={Av Tabelião Luiz Nogueira, s/n - Santo Antônio},
            city={Tianguá},
            postcode={62324-075}, 
            state={Ceará},
            country={Brazil}}

\ead{carlosalex.phys@gmail.com}





\begin{abstract}

In this paper, we address the nature of spacetime in quantum gravity in light of a new version of the holographic principle that has established a relationship between string theory and polymer holonomy structures similar to Loop Quantum Cosmology spin networks.
In front of the results found out, it is possible to argue that, for such a relationship to work, spacetime must be seen as emergent from a fundamental structure whose degrees of freedom correspond to quantum correlations only.




\end{abstract}


\maketitle

\section{Introduction}



The road to a quantum description of the gravitational phenomena has been a tough challenge.
In this sense, the main approaches to quantum gravity, superstring theory \cite{Becker:2007zj}, and loop quantum gravity (LQG) \cite{Rovelli:1988, Rovelli:2007uwt}, draw distinct scenarios for the description of the universe by taking different ways to define the fundamental elements of physical reality, including spacetime itself as a fundamental element of it.

In this way, in string theory, the fundamental objects consist of strings and branes vibrating throughout a pre-established classical spacetime. Particularly, in this context, the gravitational force is depicted by the excitations of closed strings. On the other hand, LQG consists of a background-independent approach where the fundamental objects consist of spin networks describing a spacetime that is quantum and dynamical. In fact, in the LQG context there is not even a consensus if spacetime survives or not to the LQG quantization methods \cite{Norton:2020eax}.
Other conflicts arise: string theory requires that spacetime must have $10$ dimensions, and needs the existence of supersymmetry. LQG, on its side, is a four-dimensional theory, and supersymmetry is not a feature of it.




On the other hand, recently a new route for quantum gravity has been paved by following the idea firstly introduced by  Sakharov \cite{Sakharov:1967pk}, and developed later by Jacobson \cite{Jacobson:1995ab} and others \cite{Padmanabhan:2009vy, Verlinde:2010hp},  that considers spacetime as an emergent (thermodynamical) phenomenon. 
In this case, on the basis of a version of the AdS/CFT correspondence, the so-called Black Hole Holographic Conjecture \cite{Tanaka:2002rb, Emparan:2002px, Gregory:2004vt, Gregory:2008rf}, it has been possible to argue that the quantum degrees of freedom of the theory, which is holographically related to string theory in the bulk, correspond to polymer holonomy structures similar to Loop Quantum Cosmology (LQC) spin networks \cite{Silva:2020bnn}.

Such a bridge between string theory and LQG has make it possibe to shed light on three important problems in quantum gravity: (1) the stumbling block of the big bang singularity in the AdS/CFT scenario \cite{Bak:2006nh, Engelhardt:2015gla};  (2) the quest for an induced positive cosmological constant on the $AdS$ bulk boundary; and (3) the issue of the quantum ambiguities related to the Barbero-Immirzi parameter in LQG \cite{Rovelli:1997na}.


In the present paper, by following the discussions introduced in \cite{Silva:2020bnn}, we shall address the nature of the fundamental degrees in quantum gravity, by addressing the role of spacetime in such a context.  In front of the results found out, it is possible to argue that spacetime must be not fundamental, but an emergent entity in the context of quantum gravity, whose fundamental degrees of freedom, from which spacetime itself must to emerge, will correspond to quantum correlations only. Not correlations among things, but only quantum correlations.




The paper is organized as follows: in section \eq{sec2}, we shall review the main aspects of the theory introduced  in \cite{Silva:2020bnn}. In section \eq{sec4}, we shall address the implications of such a theory for the role of spacetime in quantum gravity. In section \eq{sec5}, we shall address the fundamental degrees of freedom of quantum gravity in the light of the theory introduced in \cite{Silva:2020bnn}.  Section \eq{conc} is devoted to conclusions and discussions. In the present letter, we shall use $c = \hslash = G = 1$.



%
%

\section{Braneworld polymer structures and closed strings.}\label{sec2}



In front of the difficulties to obtain a quantum description of the gravitational phenomena, the holographic principle appears as a possible guide
\cite{'tHooft:1993gx, Susskind:1994vu}. Such a principle states that gravitational physics in $D$ spacetime
dimensions must emerge from a quantum theory living in one spatial dimension lower.

In its more successful form, the holographic principle appears as the so-called AdS/CFT conjecture \cite{Maldacena:1997re}, which corresponds to a non-perturbative version of string theory, establishing a correspondence between a type IIB string theory living in a $AdS_{5} \times S^{5}$ bulk and the quantum physics of strongly correlated many-body systems, living on the bulk boundary, described by a $U(N)$ matrix theory.  The $AdS/CFT$ correspondence has found a lot of applications in physics and cosmology, consisting in one of the most powerfull tools in quantum gravity research.


An interesting version of such a correspondence has been proposed by Emparan, Kaloper, and Fabri \cite{ Emparan:2002px}, and independently by Tanaka \cite{Tanaka:2002rb}. It has been named by Gregory as the Black Hole Holographic Conjecture (BHHC) \cite{Gregory:2008rf}. According to such a proposal, the AdS/CFT correspondence must be modified in a way that a classical gravitational theory living in a $AdS_{5}\times S^{5}$ bulk, described by string theory, is holographically related not to a quantum field theory, but to a quantum field theory coupled to gravity, i.e., a  semiclassical gravitational theory living on the bulk boundary. In this way, the BHHC brings us the novelty that gravitational degrees of freedom must be includded in the description of the boundary theory.

By following such ideas, an extension of the BHHC has been proposed recently \cite{Silva:2020bnn}. It takes into account the idea introduced by Sakharov \cite{Sakharov:1967pk}, and developed  by Jacobson \cite{Jacobson:1995ab} and others \cite{Padmanabhan:2009vy, Verlinde:2010hp} that spacetime must be seen as an emergent (thermodynamical) phenomenon. In according to such an approach, in a scenario where the AdS boundary corresponds,  in the semiclassical limit, to a four-dimensional flat FLRW universe living on an RSII brane \cite{Silva:2020bnn}, string theory in the AdS bulk must be holographically related  not to a semiclassical gravitational theory but to a quantum gravity theory, specifically to a quantum cosmological theory.

In this case, by following the discussions by Singh and Sony in \cite{Singh:2015jus}, it was possible to argue that the Hamiltonian operator describing the fundamental degrees of freedom of such a quantum cosmological theory corresponds to

\begin{eqnarray}
\mathcal{H} &=& - \frac{3V}{8\pi\xi^{2}}\sin^{2}(p\sqrt{\Delta}/2) \nonumber \\
                    &=& - \frac{3V}{16\pi\xi^{2}}[1 - e^{i p\sqrt{\Delta}}  -  e^{-i p\sqrt{\Delta}}]. \label{lqc-hamiltonian}
\end{eqnarray}

\noindent The  Hamiltonian \eq{lqc-hamiltonian} is similar to that appears in the context of LQC, with the parameter $\Delta$ corresponding to an area gap \cite{Singh:2015jus, Saini:2018tto}. Moreover, $V = a^{3}$ represents the physical volume of a fiducial
cell whose coordinate volume is chosen to be unity, with $a$ giving the scale factor in the FLRW brane. The parameter $\xi$ is given by 

\begin{equation}
\xi = (3/(16\pi \sigma))^{1/2}, \label{xi-sigma}
\end{equation}

\noindent where $\sigma$ is the brane tension.

In order to quantize the theory introduced in \cite{Silva:2020bnn}, we have that, due the form of the Hamiltonian \eq{lqc-hamiltonian}, the most natural way to do it is througth the use of polymer quantization \cite{Corichi:2007tf}. In this way, by using such a procedure, the Hamiltonian \eq{lqc-hamiltonian} can be promoted to a quantum operator:

\begin{equation}
\widehat{\mathcal{H}}  = \frac{-3V}{32\pi  \xi^2}[2\mathbb{I} - \hat{h}_{+} -  \hat{h}_{-}] \;\;, \label{q-lqc-hamiltonian}
\end{equation}


\noindent where $\mathbb{I}$ gives the identity matrix.


The basic observables in the theory introduced in \cite{Silva:2020bnn} consist of  the operators $\hat{h}_{\pm}$ in the Hamiltonian \eq{q-lqc-hamiltonian}. Such operators are holonomies, which act as $U(1)$ transformations creating the braneworld quantum gravitational states $\psi_{x_{n}} = e^{ipx_{n}}$ as:

\begin{equation}
\hat{h}_{\pm}\psi_{x} = e^{\pm i\sqrt{\Delta} p}e^{i px} = e^{i(x \pm \sqrt{\Delta}) p} = \psi_{x\pm \sqrt{\Delta}}\;\;, \label{v-equation} 
\end{equation}

\noindent where

\begin{equation}
p = - \frac{1}{4\pi}H
\end{equation}

\noindent is proportional to the Hubble rate $H$,  corresponding to the conjugate momentum to $V$ \cite{Singh:2015jus}.
Moreover, the quantum of area in the microscopic theory \cite{Silva:2020bnn} is given by   \footnote{In the Eq.\eq{xi-sigma}, we have corrected by a factor $4$ the expression given to $\Delta$ in the reference \cite{Silva:2020bnn}.
Such correction comes from the fact that in the reference \cite{Singh:2015jus}, the universe critical density in the Friedmann equations consistent with the Hamiltonian \eq{q-lqc-hamiltonian} appears multiplied by $4$. }

\begin{equation}
\Delta  = \frac{12\pi}{\sigma} \label{r-delta}.
\end{equation}


In this scenario, the holonomies appearing in the Hamiltonian \eq{q-lqc-hamiltonian} build the brane as a regular lattice:

\begin{equation}
\gamma_{\sqrt{\Delta}} = \{x \in \mathbb{R}  \mid x = n\sqrt{\Delta}, \forall n \in \mathbb{Z} \}\;\;. \label{lattice}
\end{equation}


\noindent It consists of the main result found out in \cite{Silva:2020bnn}, where the graph above corresponds to a $U(1)$ polymer structure, similar to LQC spin networks \cite{Mielczarek:2012pf}, but with the parameter $\Delta$ defined in terms of the brane tension.

In this way, an RSII brane describing a flat FLRW universe turns out to be seen as a $U(1)$ polymer quantum structure when one considers the microscopical description of its gravitational degrees of freedom.

One of the main implications of this result consists of the fact that the discreteness in the variable $x$ above, with discreteness parameter $\sqrt{\Delta}$, implies in superselection rules for the brane gravitational sector.
Since branes sources gravitationally the $AdS_{5}\times S^{5}$ spacetime by emitting or absorbing closed strings, whose couplings, $g_{s}$, can be related to the brane tension as $g_{s} \sim 1/\sigma$ \cite{Becker:2007zj}, one finds out from the Eq. \eq{r-delta},

\begin{equation}
96\pi^{4}\alpha'^{2}g_{s}  = \frac{12\pi}{\sigma} = \Delta \;, \label{delta-g}
\end{equation}

\noindent where $\alpha'$ is the Regge slope parameter.




The central result given by Eq. \eq{delta-g} reveals the relationship between two fundamental parameters belonging to different theories connected through holography.
In this way, closed string states, characterized by the string coupling $g_{s}$,  in the $AdS_{5}\times S^{5}$ bulk, are connected through holography to quantum states in the microscopic theory from which they emerge, which in their turn, are characterized by the parameter $\Delta$.

In this way, the scenario introduced in \cite{Silva:2020bnn} have made it possible to extend the BHHC in a way that a type IIB string theory living in a $AdS_{5}\times S^{5}$ manifold turn to be holographically related not to a semiclassical gravitational theory, but to a quantum gravity theory described by $U(1)$ polymer structures whose quantum degrees of freedom consists of quantum area states.

Such a version of the holographic principle have been useful to address some important problems related to quantum gravity as the stumbling block of the big bang singularity in the AdS/CFT correspondence \cite{Bak:2006nh, Engelhardt:2015gla}, and the quantum ambiguities related to the Barbero-Immirzi parameter \cite{Rovelli:1997na}. 
However, possibily the most appealing consequence of the theory introduced in \cite{Silva:2020bnn} is that it was possible to obtain a positive cosmological constant induced on the brane by quantum effects in a string theoretical scenario, which brings us a more realistic scenario in order to describe our accelerated expanding universe, enabling the contact with observational data.

\section{An interlude to the emergence of spacetime} \label{sec4}

The results introduced in \cite{Silva:2020bnn} bring us the interesting lesson that, by taking into acount the microscopic structure of branes, such objects must be seen as polymer holonomy quantum structures.

However, before to lead such results further, one must at first take care on the fact that LQG quantization methods have been used to get them. 
In this case, by considering the modern interpretations of such methods,  it has been pointed that LQG quantization technics would lead not only to the discretization of spacetime, but to the modification of its underlying structure \cite{Bojowald:2011hd}.

However, we note that, in the context of the theory introduced in \cite{Silva:2020bnn}, the geometry of the bulk boundary would be modified by quantum effects in a way that we would no longer have an $AdS$ scenario anymore. In this case, the ideas introduced in \cite{Silva:2020bnn} could find a dead end.

In order to getting out such a hard place, it is important to highlight that there is not a consensus in the LQG literature about the implications of the use of its quantization methods (for a broad discussion about this issue, and references, see \cite{Norton:2020eax}).

In this way, a more radical possibility that has been presented in the literature is that spacetime itself must not survive to LQG quantization technics, but that such methods must dissolve it into a pre-geometrical regime from which spacetime itself outgh to emerge. 
In this case, the results found out in \cite{Silva:2020bnn} can match an idea that has get a lot of attention in the context of the $AdS/CFT$ correspondence and other quantum gravity approaches \cite{Ryu:2006bv, Ryu:2006ef, Nishioka:2009un, Takayanagi:2012kg, VanRaamsdonk:2009ar, VanRaamsdonk:2010pw, Girelli:2005ii}: that spacetime must be seen as not fundamental but as an emergent phenomenon.

In the context of such discussions, it has been proposed that spacetime must emerge from fundamental elements consisting only of quantum correlations among quantum reference frames \cite{Aharonov:1984zz, Rovelli:1990pi, Toller:1996ki, Girelli:2005ii,   Bartlett:2007zz, Giacomini:2017zju}.
In such a relational approach, if we have the right pattern of correlations, the system can be cleaved into parts that can be identified as different portions of spacetime. The degree of entanglement among such parts defines the notion of spatial distance: the closer two portions of spacetime are, the greater the entanglement between them.

The astonishing notion of quantum reference frames arises in this context as related to the universality of the theory: both observers and observed must satisfy the same natural laws. In this case, if the quantum description of the world is universal, one would expect that reference frames (observers) should obey the laws of quantum mechanics, corresponding in this way to quantum systems, as a quantum particle, for exemple. 

The introduction of quantum reference frames bring us interesting questions like:  how would be the world seen by a quantum system like a quantum particle? In this way,  some interesting challenges are placed before us. For example, it has been argued that even some fundamental ideas related to our conception about physical reality, as the notion of causality must be changed, becoming relative in a scenario where one takes into account the description of the world as that given by  quantum reference frames \cite{Castro-Ruiz:2019nnl}. 

However, the situation must become even worse when one faces the problem of quantum gravity. It is because,  by considering regimes where the quantum bahavior of the gravitational phenomena becomes important,  it is necessary to conceive a quantum reference frame that obeys not only the laws of quantum mechanics, but the laws of quantum gravity itself. The point is that we do not know what these laws would be.

On the other hand, an interesting tip arises from string theory. Specifically, from the fact that, in the context of such an approach, and others related to it,  such as supergravity theories, a brane is a physical object that generalizes the notion of a point particle. 
In this case, a brane consist of a physical system that could  appear as a natural choice to a quantum reference frame in order to describe the quantum nature of gravity. 

In fact, branes consist of very interesting objects to address the issue of spacetime in a quantum gravitational context. In this sense, one of the most astonishing aspect of such objects is that,  in the context of string theory, branes can be seen as real boundaries to spacetime, with the other side being null and void of it. In this case, a brane can be seen as the edge of spacetime physics when one considers the string theoretical description of the world \cite{Barcelo:2000ta, Barcelo:2000re, Rodrigo:2007zv}.

In such a context, the results found out in \cite{Silva:2020bnn} offers an interesting framework to discuss spacetime emergence since they are deeply rooted in the idea that spacetime corresponds to a thermodynamical phenomenon. 
Moreover, in the context of the results introduced in  \cite{Silva:2020bnn}, branes appear as central objects by being identified as polymer quantum holonomy structures.

\section{Beyond spacetime: branes, quantum correlations and abstract spin networks} \label{sec5}

The result found out in \cite{Silva:2020bnn} that branes must be seen as quantum polymer structures can be used to consolidade the idea that such objects can be conceived as quantum gravitational reference frames. Its is because, due to the $U(1)$ holonomy structure assigned to such objects in such a context,  branes can be conceived as quantum clocks, i.e., temporal quantum reference frames  \cite{Girelli:2005ii, Castro-Ruiz:2019nnl, Smith:2020zms}.

However, before to take such a route, it is interesting to note that possibly a more disruptive lesson can be introduced by the ideas coming from the results of \cite{Silva:2020bnn}. In this way,  by taking branes as quantum holonomy structures, one can make the parsimonious choice of  pick up the same holonomies used to establish branes as quantum reference frames to be used also to weave the entanglement network between them.

Such a possibility arises from the fact that holonomies consist of unitary transformations connecting two quantum gravitational states, belonging (in general) to two different Hilbert spaces, through quantum entanglement \cite{Baytas:2018wjd, Bianchi:2018fmq, Mielczarek:2018jsh}:

\begin{equation}
\hat{h}_{ij} \; : \; \mathcal{H}_{i} \rightarrow \mathcal{H}_{j}\;, \label{holonomy-entanglement-1}
\end{equation}

\noindent where

\begin{equation}
\hat{h}_{ij}\hat{h}_{ij}^{\dagger}  = \mathbb{I} \;. \label{holonomy-entanglement-2}
\end{equation}

\noindent In this case, it is possible to conceive a scenario where the matrices $\hat{h}_{ij}$ carry the entire combinatorial information related to the deeper structure from which spacetime must emerge.

However, an important detail arises: that in such a scenario there will be no fundamental distinction between the holonomies describing the quantum reference frames and the holonomies describing the quantum correlations. In other words, there will be no distinction between quantum correlations and the objects correlated by them.

In fact, even though one can label some holonomies (as $h_{ii}$s, for example) to be frozen to fix the brane quantum reference frames, and others holonomies (as $h_{ij}$s ($i \neq j$), for example) to describe the connections between them, only quantum correlations are laid on the table as the fundamental elements of the theory. Not correlations among things but only quantum correlations are fundamental since even the objects we have called quantum reference frames (the correlate) must be understood as quantum correlations to their core.

It introduces a pure relational scenario where the conceptions of clocks or rods, to measure distances or time intervals, and even the conception of objects to be localized in spacetime, becomes obsolete. In this way,  no geometrical, nor topological information will make sense anymore, but only combinatorial information remains.

It is possible to understand such a pure combinatorial scenario revealed by branes, by noting that if one simply labels the $\mathcal{H}_{i}$s, in the Eqs \eq{holonomy-entanglement-1} and \eq{holonomy-entanglement-2},  as the Hilbert spaces of $N$ branes, the indices $i,j$ above will start to range as $i,j = 1,...,N$.
In this way,  the $\hat{h}_{ij}$s will be generalized to $U(N)$ unitary matrices by  describing both the quantum reference frames and the quantum correlations between them.

By relying on such a $U(N)$ algebraic structure, it is possible to write down the theory introduced above on a basis that fits its pure combinatorial characterization. 
In this way, due to the interesting fact that a vector in the representation space, $\textit{\textbf{u}}(N)$, of the unitary group corresponds to an intertwiner with $N$ legs dressed by $SU(2)$ representations \cite{Girelli:2005ii}, 
one can write the $U(N)$ group elements $\hat{h}_{ij}$ in terms of $\textit{\textbf{u}}(N)$ basis vectors as

\begin{equation}
 \hat{h}_{ij} = e^{M_{ij}} = \mathbb{I} + M_{ij} + \frac{1}{2}M_{ij}^{2} +  ...\;\;\;,  \label{spin-net-exp}
\end{equation}

\noindent where $M_{ij} \in \textit{\textbf{u}}(N)$ corresponds to a $N$-leg intertwiner.

Such intertwiners consists of abstract spin networks \cite{Girelli:2005ii}, which differ from the usual LQG ones in the fact that they do not describe a quantized spacetime, but consist of graphs labeled by $SU(2)$ representations carrying only combinatorial information, without any reference to a background geometry or topology \cite{Girelli:2005ii}. In this way, such structures fit the scenario where
only quantum correlations exist, and the degrees of freedom are purely algebraic and combinatorial.

Consequently, in contrast with the theory introduced in \cite{Silva:2020bnn}, the degrees of freedom  related to quantum gravity, in a more fundamental level, must not consist in quantum area states, nor in closed string states, but in purely combinatorial degrees of freedom. In this case, both closed string states and quantum area states  must emerge from the  $SU(2)$ combinatorial rules governing the angular momentum eigenvalues attached to the intertwiner legs.

For example, by embedding the intertwiners into a manifold, one can trace a way the quanta of area \eq{r-delta} will be written in terms of angular momentum eigenvalues, by the choice of some area spectrum. Such a procedure can be used to understand how the geometrical features of such a manifold can be related to the combinatorial rules governing the intertwiners, in a similar way we have in usual LQG \cite{Liko:2005kb}. Moreover, the way branes will source the $AdS$ spacetime must be affected by the $SU(2)$ combinatorial rules, which must have consequences to the spectrum of closed strings in the $AdS$ bulk.

Following such discussions, it is possible reformulate the conjecture introduced in \cite{Silva:2020bnn}, relating string theory to polymer structures, in the following way:  a $AdS_{5} \times S^{5}$ bulk, containing a type IIB string theory, as well as its boundary,  must emerge from a $U(N)$ matrix theory, described by abstract spin networks, whose fundamental degrees of freedom are quantum correlations without correlate. Such an emergence must occur in the large-$N$ limit, where in the present context, the large-$N$ limit will correspond to that where the number of quantum correlations becomes large.

\section{Conclusions and discussions.}\label{conc}

In order to reconcile the results found out in \cite{Silva:2020bnn} with modern perspectives on LQG quantization methods, it becomes necessary to make use of the idea that spacetime must be an emergent entity in our description of physical reality. 
It agrees with the core of the argument used in \cite{Silva:2020bnn} which is based on the idea that spacetime is a thermodynamical phenomenon.

In this way, the results introduced in the present article trace a scenario where spacetime must emerge from a pre-geometric regime governed by a $U(N)$ matrix theory written in terms of abstract spin networks describing quantum correlations without correlate.
In such a scenario, the fundamental degrees of freedom are not geometrical in any sense, but they are purely combinatorial.

It is interesting to note that, by considering the present discussions,  it will be possible to choose a structure of quantum reference frames from which spacetime will emerge in a non-unique way. In fact, it will be a large number of ways to make such a kind of choice, depending on the holonomies we choose to label as describing the quantum reference frames. It could shed some light on the so-called bulk locality paradox, which consists of the fact that, in the AdS/CFT formulation of the holographic principle, there is not a unique way a bulk operator can be reconstructed from boundary ones \cite{Almheiri:2014lwa, Pastawski:2015qua}.

Moreover, the results introduced in the present letter, may change our perspective about some aspects of the AdS/CFT correspondence as an approach based on string theory. It is because,  by seeing such a theory as emergent, some peculiar features belonging to it, like extra dimensions or supersymmetry, do not need to have a microscopic counterpart described by spin networks. But they can be seen as emergent phenomena. Several papers in the literature have supported the idea that extra dimensions must emerge from a fundamental four-dimensional quantum gravity theory, starting from the seminal work by Arkani - Hamed, Cohen, and Georgi \cite{Arkani-Hamed:2001kyx}. Moreover, examples of emergent spacetime supersymmetry have been found in the context of condensed matter physics \cite{Grover:2013rc}, where spacetime SUSY naturally emerges at low energy and at long distances, although
the microscopic ingredients of the theory are not supersymmetric.

However, the main contribution of our results is that only quantum correlations must exist as the elements of physical reality. Not correlations among things, but only quantum correlations. It is based on a key detail about the approach introduced in the present letter: the holonomies describing quantum reference frames have no intrinsic property that could distinguish them from the holonomies describing quantum correlations.
Consequently, even what we have called quantum reference frames (the correlate) can be understood as quantum correlations, to the core. We note that the construction of the $U(N)$ matrix theory describing the pre-geometric regime from which spacetime must emerge is rooted in such a fact.

Even though to conceive a mechanism to spacetime emergence in the context of quantum gravity remains a chalenge, the  ideas introduced in the present article can promote a change of route in the discussions about spacetime emergence in this context, by shedding some light on some important difficults related to this issue.
 
For example, when one considers  the usual form of the holographic principle, as we have in the AdS/CFT context, it is considered that spacetime geometry must emerge holographically from a quantum theory living in a spatial dimension lower.
However, some authors have pointed out a conceptual problem in such a construction. It is because, in the situation where the bulk emerge from the boundary, the last should to be more fundamental than the first. It is dificult to reconcile with the fact that, in the AdS/CFT scenario, the duality between the boundary and the bulk theories is exact, and thus symmetrical.  

Such a kind of conceptual barrier is linked with the fact that it is difficult to completely leave the idea of the existence of spacetime in such a context, and in this way, we still need the existence of a locus (the boundary) from which the bulk itself will emerge. In fact, this is rooted in some deep questions that still haunt the issue of spacetime emergence: how could physics exist beyond spacetime, and how could things exist, and become entangled, without some loci where and when they happen and change?

A similar dificult is shared by  LQG where there is a lack of consensus if spacetime survives or not to the quantization methods used in this approach \cite{Norton:2020eax}, and if spin networks can exist or not beyond spacetime. In fact, how could spin networks carry combinatorial information without referring to objects to be combined and, as a matter of course, to a locus in which such objects live?

However, a possible way to have combinatorial information without the necessity of a locus for the existence of objects is that, in fact, localized objects do not exist anyway, and only correlations without correlate correspond to the elements of physical reality \cite{Norton:2020eax}. Such a point of view has been launched in the context of the so-called Structural Realism \cite{Ladyman:1998}.

In this sense, the results introduced in the present paper require a new perspective about the cosmos, when one consider its quantum properties, which must differ from that provided by the usual Copenhagen interpretation. In fact, the ideas introduced in the present paper are close to those proposed by Mermin in the so-called Ithaka interpretation of quantum mechanics \cite{Mermin:1996mr, Mermin:1998cg}. According to such an interpretation, only quantum correlations must exist. Not correlations among things, but only quantum correlations. However, in a different way from Mermin's point of view, in the approach introduced in the present paper, quantum correlations must not exist in spacetime, but they must precede spacetime itself.

\section*{Acknowledgements}
The author acknowledges F.G. Costa, M.B. Cruz, G.Alencar, R.R. Landim, C. A.S. Almeida, R.V. Maluf, J.E.G. Silva,  F. A. Brito and I. C. Jardim for the reading of previous versions of the manuscript and useful discussions. 

\end{document}